\documentclass[aps,floatfix,prd,twocolumn,a4paper,10pt,citeautoscript,
preprintnumbers,superscriptaddress,longbibliography,showpacs]{revtex4-1} 

\usepackage[english]{babel}	
\usepackage{changes}
\usepackage{amsmath}		
\usepackage{amssymb}		
\usepackage{bm}				
\usepackage{bbm}			
\usepackage{grffile}        
\usepackage{graphicx}		
\usepackage{graphics}		

\DeclareGraphicsExtensions{.jpg,.png,.pdf,.eps}	
\usepackage[colorlinks=true, pdfstartview=FitV,linkcolor=blue, citecolor=blue, urlcolor=blue]{hyperref}

\newcommand{\Figref}[1]{Fig.~\ref{#1}}
\newcommand{\Eqref}[1]{\eqref{#1}}
\newcommand{\groupZ}[1]{\mathbb{Z}_{#1}}	
\newcommand{\Exp}[1]{\text{e}^{#1}}
\renewcommand\Re{\mathrm{Re}}
\renewcommand\Im{\mathrm{Im}}

\newcommand{\F}{\mathcal{F}}

\begin{document}
\title{Phase diagram of dirty two-band superconductors and observability 
of impurity-induced \texorpdfstring{$s+is$}{s+is} state}
\author{Mihail~Silaev}		
\affiliation{Department of Theoretical Physics, 
KTH-Royal Institute of Technology, Stockholm, SE-10691 Sweden}
\affiliation{Department of Physics and Nanoscience Center, 
University of Jyv\"askyl\"a, P.O. Box 35 (YFL), FI-40014 
University of Jyv\"askyl\"a, Finland}
\author{Julien~Garaud}		
\affiliation{Department of Theoretical Physics, 
KTH-Royal Institute of Technology, Stockholm, SE-10691 Sweden}
\author{Egor~Babaev}			
\affiliation{Department of Theoretical Physics, 
KTH-Royal Institute of Technology, Stockholm, SE-10691 Sweden}

\begin{abstract}

We investigate the phase diagram of dirty two-band superconductors. 
This paper primarily focuses on the properties and observability of the 
time-reversal symmetry-breaking $s+is$ superconducting states, which 
can be generated in two-band superconductors by interband impurity 
scattering. We show that such states can appear in two distinct ways. 
First, according to a previously discussed scenario, the $s+is$ state can 
form as an intermediate phase at the impurity-driven crossover between 
$s_{\pm}$ and $s_{++}$ states. We show that there is a second scenario 
where  domains of the $s+is$ state exists in the form of an isolated 
dome inside the $s_{\pm}$ domain, completely detached from the transition  
between $s_{\pm}$ and $s_{++}$ states. We demonstrate that in both cases 
the $s+is$ state, generated by impurity scattering exists in an extremely 
small interval of impurity concentrations. Although this likely precludes 
direct experimental observation of the $s+is$ state formation due to this 
mechanism, this physics leads to the appearance of a region inside both 
the $s_{\pm}$ and $s_{++}$  domains with unusual properties due to 
softening of normal modes.  
\end{abstract}
\date{\today}
\maketitle

 \section{Introduction}
 
The physics of unconventional superconductors with multiple broken symmetries 
has attracted a lot of attention for a long time \cite{Sigrist.Ueda:91}. 
Of particular interest is the Broken-Time Reversal Symmetry state (BTRSs) 
which has been proposed to exist in Sr$_2$RuO$_4$\cite{Mackenzie.Maeno:03,
Huang.Scaffidi.ea:16} and heavy-fermion compounds \cite{Joynt.Taillefer:02}. 
In these systems the discussed BTRSs, is commonly referred to as the $p+ip$ 
state by analogy with the chiral A-phase of superfluid $^3$He confined in a 
thin slab \cite{Volovik:88,Sigrist.Ueda:91}. There, the BTRSs appears in 
combination with parity breaking.

In multiband systems, a different type of time-reversal symmetry-breaking 
state, termed $s+is$, can appear. It is fundamentally different from the 
well studied $p+ip$ state, and its principal distinction is that it does 
not break any crystal symmetries and thus represents a type of superconducting 
state beyond the lattice point group-based classification. Such states 
were discussed in a wide range of systems, and in particular in three-band 
superconductors with frustrated interband interactions \cite{Ng.Nagaosa:09,
Stanev.Tesanovic:10,Maiti.Chubukov:13}.

The $s+is$ states have been predicted to host a broad range of interesting 
new phenomena, among which can be mentioned different massless \cite{Lin.Hu:12} 
and ``phase-density mixed" \cite{Carlstrom.Garaud.ea:11a,
Stanev:12,Maiti.Chubukov:13,Marciani.Fanfarillo.ea:13} collective modes, 
unconventional thermoelectric properties \cite{Silaev.Garaud.ea:15,
Garaud.Silaev.ea:16}, additional mechanisms of damping of the vortex motion 
\cite{Silaev.Babaev:13} and unconventional magnetic signatures induced by 
defects \cite{Maiti.Sigrist.ea:15,Lin.Maiti.ea:16}. Multiple broken symmetries 
in $s+is$ superconductors give rise to several strongly disparate coherence 
lengths. This can lead to a state with attractive intervortex interaction 
originating in the magnetic field penetration length being smaller than some, 
and larger than other coherence lengths \cite{Carlstrom.Garaud.ea:11a,
Garaud.Silaev.ea:16a}. Besides vortices, the $s+is$ state also allows other 
types of topological excitations that include domain walls and skyrmions 
\cite{Garaud.Carlstrom.ea:11,Garaud.Carlstrom.ea:13,Garaud.Babaev:14}. 
The $s+is$ state also exhibits complex beyond-mean-field physics with new 
fluctuation-induced phases \cite{Bojesen.Babaev.ea:13,Bojesen.Babaev.ea:14,
Carlstroem.Babaev:15,Hinojosa.Fernandes.ea:14}.

The $s+is$ state can have various microscopic physics origins. 
It was recently shown to be generated by quasiparticle scattering in 
multiband superconductors with repulsive interaction of electrons in 
different bands \cite{Bobkov.Bobkova:11,Stanev.Koshelev:14}.  
This mechanism would be rather generic for iron-pnictide superconductors, 
where the pairing is generated by the interband electron-electron repulsion
\cite{Hirschfeld.Korshunov.ea:11}, producing the so-called $s_\pm$ 
superconducting state with a sign change between the order parameter 
components in different bands \cite{Mazin.Singh.ea:08,Chubukov.Efremov.ea:08}.  
That is, for two components of order parameter $|\Delta_i|e^{i\varphi_i}$, 
the state has  $\varphi_1=\varphi_2 +\pi$ in contrast to the $s_{++}$ state 
where $\varphi_1=\varphi_2$.
  
Existence of $s+is$ state localized near the surface of a two-band $s_{\pm}$ 
superconductor has been investigated in \cite{Bobkov.Bobkova:11}, while it 
was later proposed that interband impurity scattering can generate $s+is$ 
state in the bulk \cite{Stanev.Koshelev:14}. It has further been argued 
that the disorder-induced transition from $s_{\pm}$ to $s_{++}$ state in 
two-band superconductors can occur in two qualitatively different ways. 
The first one is a crossover without additional symmetry breaking when 
the superconducting gap in one of the bands crosses zero, as a function of 
impurity concentration \cite{Efremov.Korshunov.ea:11}. The second possible 
scenario involves $s_{\pm}$ to $s_{++}$ state transformation through the 
intermediate complex $s+is$ state $\varphi_1=\varphi_2 +\delta$ with 
$\delta \ne 0, \pi$ \cite{Stanev.Koshelev:14}.     
 
In this paper we calculate phase diagrams of superconducting states in 
the presence of interband scattering, and discuss the basic properties 
of emerging BTRSs. We point out that in general the phase diagrams are 
quantitatively and also qualitatively different from the one sketched 
in Ref.~\onlinecite{Stanev.Koshelev:14}. In particular we show that a 
domain of the $s+is$ state is not necessarily attached to $s_{\pm}$ 
to $s_{++}$ crossover line but under certain conditions arises inside 
the $s_{\pm}$ state. 
We further analyze, within the framework of microscopic Ginzburg-Landau 
expansion, the properties of the transition lines to the $s+is$ state. 
Finally, we conclude that in the superconductors described by a 
weak-coupling two-band theory, $s+is$ occupies a very narrow region of 
the phase diagram as a function of impurity concentration and temperature.
Thus, in such two-band systems, it is extremely unlikely to observe the 
impurity-induced $s+is$ states. However, its presence on the phase diagram 
can influence the properties of the superconducting state in a wider region 
of the phase diagram even outside the $s+is$ domain, as we discuss below. 

The rest of this paper is organized as follows. Section \ref{Sec:Microscopic} 
introduces the basic framework and briefly discusses the known properties 
of impurity driven $s_{\pm}$ to $s_{++}$ crossover in two-band superconductors. 
Next, Section \ref{Sec:PhaseDiagrams} presents the numerically calculated 
phase diagrams, in several characteristic cases. Section \ref{Sec:Analytics} 
addresses the general mean-field properties of the $s+is$ transition regions. 
The mean-field critical behavior at the $s_{\pm}/s+is$ and $s_{++}/s+is$ 
transition lines is discussed in terms of the Ginzburg-Landau theory, 
in Section \ref{Sec:GL}, and finally our conclusions are given in Section
\ref{Sec:Conclusion}.  

\section{Microscopic model}
\label{Sec:Microscopic}

We consider a superconductor with two overlapping bands at the Fermi level. 
Within the quasiclassical approximation, the band parameters characterizing 
the two different sheets are the partial densities of states (DOS) $n_k$, 
labelled by the band index $k = 1, 2$. The order parameter is determined, 
using a microscopic theory formulated in terms of quasiclassical propagators  
$g_k$ and $f_k$, the normal and anomalous Green's functions obeying the 
normalization condition $|f_k|^2 + g_k^2 =1$. The system of Eilenberger 
equations for a spatially homogeneous two-band superconductor with 
impurities reads as \cite{Gurevich:03}: 
\begin{align} 
  \omega_n f_1 &= \Delta_1 g_1 + \gamma_{12} ( g_1f_2 - g_2f_1) 
\label{Eq:Eilenberger:1}  \\
  \omega_n f_2 &= \Delta_2 g_2 + \gamma_{21} ( g_2f_1 - g_1f_2) 
\label{Eq:Eilenberger:2}  \,,
\end{align}
where $\omega_n =( 2n+1)\pi T $, $n\in \groupZ{}$ are the fermionic 
Matsubara frequencies, $T$ is the temperature and $\gamma_{kk^\prime}$ 
are the interband scattering rates proportional to the impurity 
concentration. 
The components of the order parameter $\Delta_k= |\Delta_k|e^{i\varphi_k}$ 
are determined by the self-consistency equation 
\begin{equation}\label{Eq:SelfConsistency1}
 \Delta_k =2\pi T  \sum_{n=0}^{N_d} 
 \sum_{\,k,k^\prime=1,2} \lambda_{kk^\prime} f_{k^\prime} (\omega_n),
\end{equation}
for the Green's function that satisfy Eqs.~\Eqref{Eq:Eilenberger:1} and
\Eqref{Eq:Eilenberger:2}. Here  $N_d=\Omega_d/(2\pi T)$ is the summation 
cut-off at Debye frequency $\Omega_d$. The diagonal elements $\lambda_{kk}$ 
of the coupling matrix $\hat\Lambda$ in the self-consistency equation 
\Eqref{Eq:SelfConsistency1}, describe the intraband pairing, while the 
interband interaction is determined by the off-diagonal terms which can 
be either positive or negative. 
In the following, we consider the latter case which corresponds to the 
interband repulsion, favoring the sign changing $s_\pm$ state. The 
interband coupling parameters and impurity scattering amplitudes satisfy 
the symmetry relation \cite{Gurevich:03}:
\begin{equation}\label{Eq:Symmetry}
 \lambda_{ij}= - \lambda_J/n_i~~\text{and}~~\gamma_{ij}= n_j\Gamma \,,
\end{equation} 
where $\lambda_J>0$ and $n_{1,2}$ are the partial densities of states 
in the two bands. 
 
In general, the $s_\pm$ state is not favoured by the impurity scattering, 
which tends to average out the order parameter over the whole Fermi surface, 
suppressing the critical temperature. Still, provided the interband 
pairing interaction is weak, superconductivity can be transformed into 
a $s_{++}$ state and survive even in the limit $\Gamma\gg T_{c0}$, 
characterized by the critical temperature $T_{c\infty}$ which reads 
as \cite{Stanev.Koshelev:14,Hirschfeld.Korshunov.ea:11}: 
\begin{equation}\label{Eq:Tc}
 \ln (T_{c0}/T_{c\infty}) =  n_1 (w_{11}+w_{12}) + n_2 (w_{22}+w_{21}) \,,
\end{equation}
where $T_{c0}$ is the critical temperature without interband scattering,
$\hat w = \hat \Lambda^{-1}- \lambda^{-1}\hat I $, and $\lambda$ is the 
maximal eigenvalue of the coupling matrix $\hat\Lambda$ with the elements 
$\lambda_{kk^\prime}$. 
According to Eq.~\Eqref{Eq:Tc}, one can see that the interband interaction 
$\lambda_J$ should be sufficiently weak, in order to avoid a drastic 
suppression of the critical temperature in the $s_{++}$ state.
To derive the criterion note that $n_1 w_{11} + n_2 w_{22} > 0$,
so that the r.h.s. of the Eq.\Eqref{Eq:Tc} is larger than 
$n_1 w_{12} + n_2 w_{21} = \lambda_J /(\lambda_{11}\lambda_{22})$. 
Therefore in order to have $T_{c\infty}$ not much smaller than 
$T_{c0}$, we require the following condition to be fulfilled:  
\begin{equation} \label{Eq:TcRestriction}
 \lambda_J /(\lambda_{11}\lambda_{22}) <1 \,.
\end{equation}  
   
Below we study the phase diagrams, given by the formalism of 
Eqs.~\Eqref{Eq:Eilenberger:1}, \Eqref{Eq:Eilenberger:2} and 
\Eqref{Eq:SelfConsistency1}, as functions of $T$ and $\Gamma$, for 
various pairing coefficients $\lambda_{kk^\prime}$. The restriction 
on pairing interactions [\Eqref{Eq:TcRestriction}] will be shown to 
set up rather strong limitations on the size of the $s+is$ domains 
as a function of the effective impurity concentration $\Gamma$.

\section{Phase diagrams}
\label{Sec:PhaseDiagrams}

\begin{figure*}[!htb]
\hbox to \linewidth{ \hss
\includegraphics[width=0.9\linewidth]{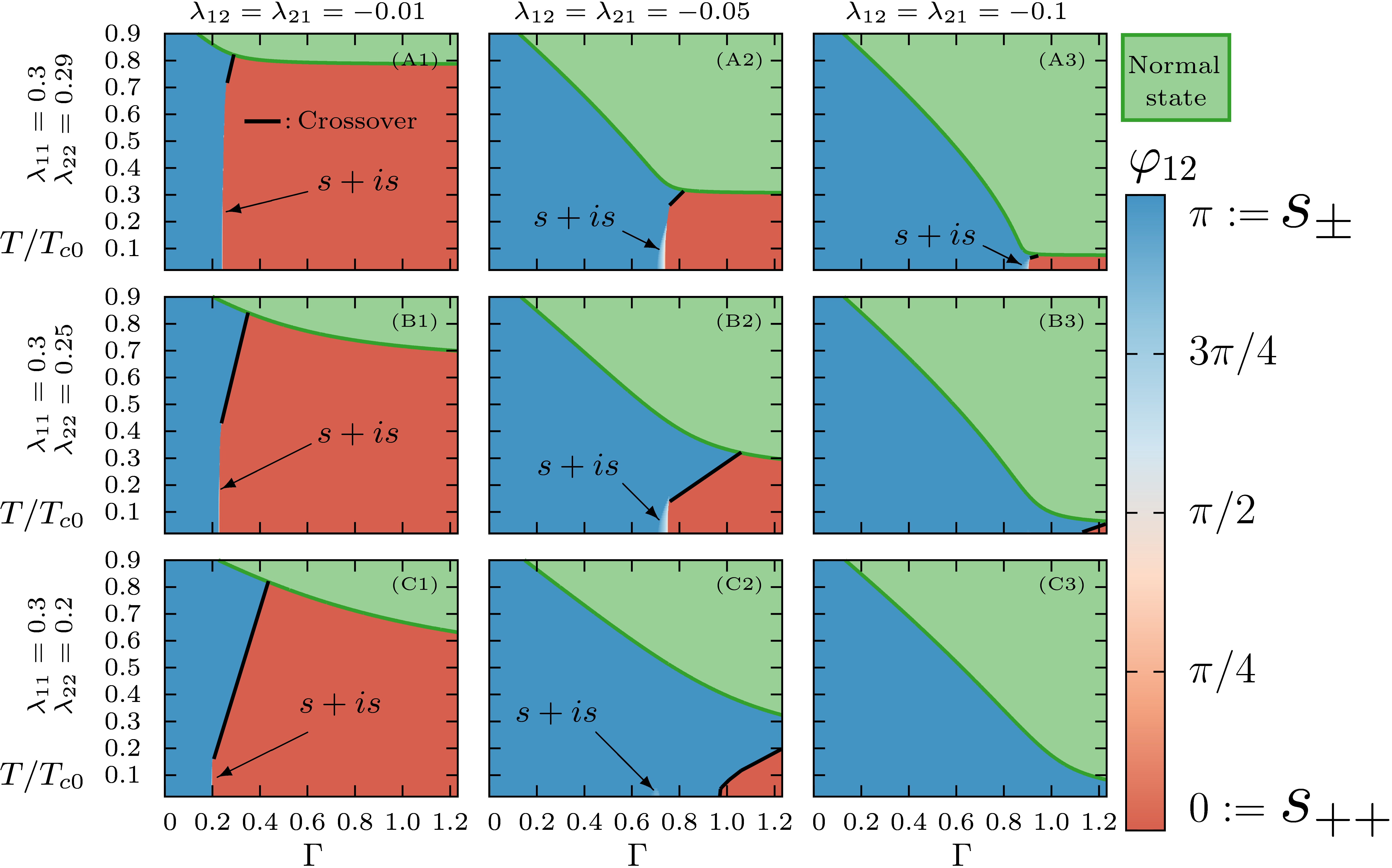}
\hss}
\caption{
(Color online) -- 
Phase diagrams of two-band superconductors with interband impurity 
scattering. These show the values of the lowest-energy-state relative 
phase $\varphi_{12}=\varphi_2-\varphi_1$ between the components of the 
order parameter, as function of temperature and interband scattering 
$\Gamma$. Different lines show different values of the intraband coupling 
parameters $\lambda_{11}$ and $\lambda_{22}$, while different columns 
correspond to different value of the interband couplings $\lambda_{12}$ 
and $\lambda_{21}$. The green solid line shows the critical temperature 
of the superconducting phase transition $T_c (\Gamma)$ and the solid 
black line shows the zero of $\Delta_2$, that is the crossover between 
$s_\pm$ and $s_{++}$ states. 
In panels A1--A3, B1, B2 and C1, the crossover line is attached to the 
$s+is$ dome. In the panel C2, the crossover line is detached from the 
$s+is$ dome, while in the panel B3 it exists without an $s+is$ dome. 
}
\label{Fig:Diagram}
\end{figure*}
We construct phase diagrams in the plane of parameters $\Gamma,T$ of a 
two-band superconductor with interband impurity scattering. For that 
purpose, we solve numerically the system of Eilenberger equations 
\Eqref{Eq:Eilenberger:1} and \Eqref{Eq:Eilenberger:2} together with the 
self-consistency equation \Eqref{Eq:SelfConsistency1}, according to the 
procedure described in the Appendix. The results are shown in 
\Figref{Fig:Diagram}, which demonstrates the role of impurities on the 
state properties, for various representative cases. The different rows in 
\Figref{Fig:Diagram}, respectively, display (A) nearly degenerate bands with 
$\lambda_{11}=0.3$, $\lambda_{22}=0.29$, (B) intermediate band disparity 
$\lambda_{11}=0.3$, $\lambda_{22}=0.25$, and (C) strong band disparity 
$\lambda_{11}=0.3$, $\lambda_{22}=0.2$. For each of these, the different 
columns show, respectively, weak (1), intermediate (2), and strong (3) 
interband pairing interactions (as compared to the intraband couplings).
 
\begin{figure*}[!htb]
\hbox to \linewidth{ \hss
 \includegraphics[width=0.9\linewidth]{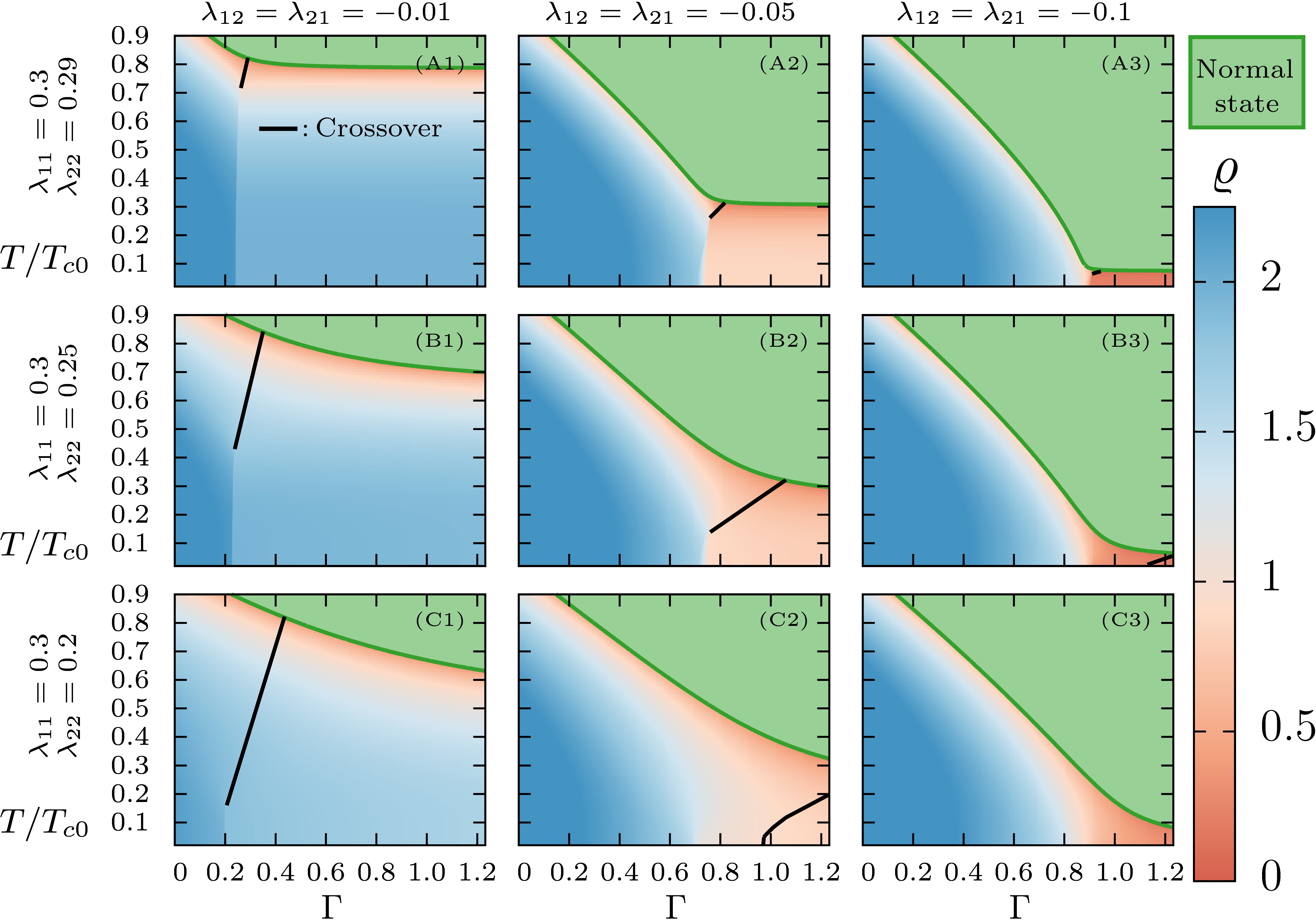}
\hss}
\caption{ 
(Color online) -- The total density $\varrho$, defined as 
$\varrho^2=|\Delta_1|^2+|\Delta_2|^2$ in two-band superconductors 
with interband impurity scattering, as a function of temperature and 
interband scattering $\Gamma$.  Different lines show different values 
of the intraband coupling parameters $\lambda_{11}$ and $\lambda_{22}$, 
while different columns correspond to different values of the interband 
couplings $\lambda_{12}$ and $\lambda_{21}$. The green solid line shows 
$T_c (\Gamma)$ and the solid black line shows the zero of $\Delta_2$, 
that is the crossover between $s_\pm$ and $s_{++}$ states.
}
\label{Fig:Density}
\end{figure*}

The crossover line between $s_{\pm}$ and $s_{++}$ states, indicated by 
the thick solid lines on the diagrams \Figref{Fig:Diagram}, is associated with 
the vanishing of $\Delta_2$ as we choose $\lambda_{11}>\lambda_{22}$. 
Everywhere else, both components of the order parameter remain finite. 
\Figref{Fig:Density} displays the total density $\varrho$, defined as 
$\varrho^2=|\Delta_1|^2+|\Delta_2|^2$, that corresponds to the various 
regimes shown in the phase diagrams \Figref{Fig:Diagram}.
The crossover generically occurs for temperatures close to $T_c(\Gamma)$ 
and at lower temperatures the $s_\pm$ and $s_{++}$ domains in the 
$(\Gamma,T)$ phase diagram may be separated by an intermediate $s+is$ 
state. When this phase is realized, it shows up as a dome that extends 
down to $T=0$. As discussed below, in Section~\ref{Sec:GL}, both the 
$s_{\pm}/s+is$ and $s_{++}/s+is$ transitions lines are of the second order, 
at the mean-field level.

The $s+is$ state is characterized by a relative phase that differs for $0$ 
or $\pi$. That possibility can be understood heuristically in the following 
way: while the interband pairing enforces a $\pi$ phase difference at zero 
impurities, the impurity scattering favors a phase difference of zero. 
As the interband pairing and impurity scattering favor different values of 
the phase locking, the system has to compromise between those behaviors and 
it can happen that the optimal phase locking is neither $0$ nor $\pi$: 
the $s+is$ state.

In general the $s+is$ dome exists at temperature $T<T_c$. However for 
nearly degenerate bands it starts very close to $T_c$ [see 
\Figref{Fig:Diagram}-(A1)]. As demonstrated in the next Section, in case 
of exact degeneracy $\lambda_{11}=\lambda_{22}$, the transition from 
$s_\pm$ to $s_{++}$ always occurs through the BTRSs domain, which 
extends to the $T=T_c(\Gamma)$ curve of the $(T,\Gamma)$ diagram. 
Increasing the disparity of intraband coupling constant $\lambda_{11}-\lambda_{22}$, 
disconnects the $s+is$ dome from the $T_{c}(\Gamma)$ curve, replacing 
it by the crossover line in a certain temperature interval. For small 
disparity $\lambda_{11}-\lambda_{22} \ll \lambda_{11}$ one can show 
that the $s+is$ domain starts at 
$T^* \propto T_c[1+\alpha(\lambda^{-1}_{11}-\lambda^{-1}_{22})]$ where 
$\alpha \sim 1$. This tendency agrees with  that shown in the first 
column of the \Figref{Fig:Diagram}. Simultaneously with the $T^*$ 
suppression, the band disparity extends the crossover line down to 
lower temperature which finally goes to $T=0$ eliminating the domain 
of the $s+is$ state attached to the $s_\pm \to s_{++}$ crossover line. 

However, in contrast to the phase diagrams reported in 
Ref.~\onlinecite{Stanev.Koshelev:14}, we show that the $s+is$ state forms
under more general conditions as an isolated dome inside the $s_{\pm}$ 
region on the phase diagram, entirely detached from the $s_\pm \to s_{++}$ 
crossover line. This effect is demonstrated in the second column of 
\Figref{Fig:Diagram}, where we take a larger value of the interband coupling 
in order to increase the width of the $s+is$ region, to make it more visible 
on the diagrams. On the other hand this set of plots demonstrates the 
general tendency governing the size of the BTRS domain which grows with 
increased interband coupling. 
At the same time, however, the critical temperature of the $s_{++}$ state 
is exponentially suppressed, according to Eq.~\Eqref{Eq:Tc}, so that 
basically the relevant values of $\lambda_{12}$ are restricted by 
Eq.~\Eqref{Eq:TcRestriction} which does not allow increasing significantly 
the size of the BTRS domain in the phase diagram. 

\section{Properties of the \texorpdfstring{$s+is$}{s+is} domain}
\label{Sec:Analytics}

\begin{figure}[!htb]
\hbox to \linewidth{ \hss
\includegraphics[width=0.75\linewidth]{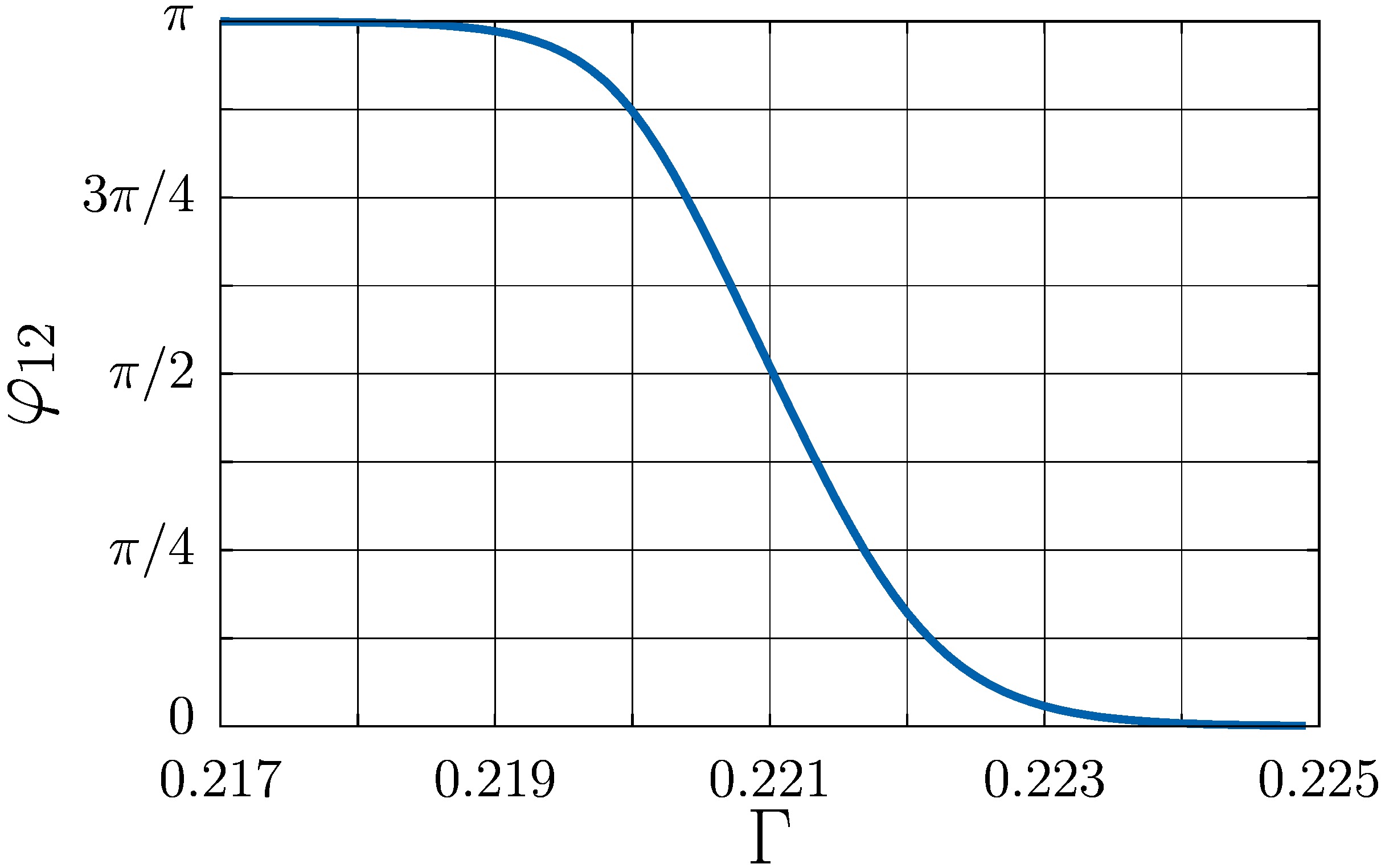}
\hss}
\caption{(Color online) -- 
This shows a close-up view of the transition from $s_\pm$ to $s_{++}$ 
with an intermediate $s+is$ phase, at $T/T_{c0}=0.3$ and for parameters 
corresponding to panel (B1) in \Figref{Fig:Diagram}. 
This demonstrates that for sufficiently low temperature, in the case of 
weak interband pairing, the $s+is$ state is realized as an intermediate 
state between $s_\pm$ and $s_{++}$ regions. However, the width of this 
region is  extremely narrow.
}
\label{Fig:Transition-zoom}
\end{figure}

The $s+is$ state is formed quite generically in case of nearly degenerate 
bands near the impurity-driven $s_{\pm}/s_{++}$ crossover. However, we find 
that it occupies only a vanishingly small region of the phase diagrams. 
In the case of weak interband pairing, defined according to 
Eq.~\Eqref{Eq:TcRestriction} as $\lambda_J < \lambda_{11}\lambda_{22}$ 
(see first column in \Figref{Fig:Diagram}), the two lines of second-order 
phase transitions $s_\pm\to s+is$ and $s+is\to s_{++}$ almost overlap. 
There is actually a very narrow region in between that requires rather 
extreme fine tuning of material parameters, as can be seen in a close-up 
view in \Figref{Fig:Transition-zoom}.
 
On the other hand, as can be seen for example in the second column of 
\Figref{Fig:Diagram}, it appears that increasing the interband interaction 
tends to widen the region where the $s+is$ state is realized. However, this growth 
is limited by the strong $T_c$ suppression at the $s_{\pm}/s_{++}$ crossover  
which sets the upper limit for $\lambda_J$ [\Eqref{Eq:TcRestriction}]. 
Hence for all possible values of interband pairing the $s+is$ domain still 
represents a vanishingly small region of the phase diagrams. Besides that, 
the $s+is$ transition lines go almost parallel to the $T$ axis, and therefore 
can be detected only by changing the effective impurity concentration, which 
would make it extremely challenging to realize the $s+is$ states
via this mechanism. 
Note also that increasing the band disparity extends the $s_{\pm}/s_{++}$ 
crossover line and restricts the impurity-induced $s+is$ state only to very 
low temperatures.  
Below, we study the shape and size of the $s+is$ domain, in the case of degenerate 
bands. The conclusions obtained are still qualitatively correct in case of 
moderate band disparities. 
    
\subsection{Size of \texorpdfstring{$s+is$}{sis} 
			transition region: the case of degenerate bands}
\label{SubSec:DegenerateBands}
 
To give an analytical estimate of the size of the $s+is$ domain, we consider 
the simplified case of identical superconducting bands $n_1=n_2$ so that 
$\gamma_{12}=\gamma_{21}=\gamma$ and $\lambda_{11}=\lambda_{22}=\lambda$. 
Under such assumptions, the superconducting gaps can be chosen in a 
symmetrical form $\Delta_1=\Delta_2^*=\Delta e^{i\varphi}$, so that 
$f_1=f_2^*=f e^{i\alpha}$, and $g_1=g_2=g$. Hence the real and imaginary 
parts of the self-consistency equation \Eqref{Eq:SelfConsistency1} become 
\begin{align} 
 2\pi T\sum_{n=0}^{N_d}\frac{g}{\omega_n} = \frac{1}{\lambda-\lambda_J}
 \,,\label{Eq:SelfConsistency21} \\ 
 2\pi T\sum_{n=0}^{N_d}\frac{g}{\omega_n+2\gamma g} = \frac{1}{\lambda+\lambda_J}
 \,.\label{Eq:SelfConsistency22}
\end{align} 
These equations are simultaneously satisfied in the $s+is$ phase, 
while $s_{++}$ and $s_{\pm}$ states are described by either 
Eq.~\Eqref{Eq:SelfConsistency21} or Eq.~\Eqref{Eq:SelfConsistency22}, 
respectively. 
 
The transition lines to the $s+is$ state can be found from the system 
Eqs.~\Eqref{Eq:SelfConsistency21} and \Eqref{Eq:SelfConsistency22}. 
First of all, we show that in the considered case of exactly degenerate 
bands, the $s+is$ domain extends up to the critical temperature. Indeed, 
at $T=T_{c}$, one can put $g=1$ so that Eqs.~\Eqref{Eq:SelfConsistency21} 
and \Eqref{Eq:SelfConsistency22} become the definitions of the critical 
temperature for $s_{++}$ and $s_{\pm}$ states, respectively. They are 
simultaneously satisfied at a single value for interband impurity 
scattering, $\gamma^*$, which determines the $s_\pm /s_{++}$ transition. 
Hence according to the Eq.~\Eqref{Eq:SelfConsistency21}, the $s+is$ state 
exists at this point when $T=T_{c}(\gamma^*)$.

For arbitrary temperatures $T<T_c$, the system of Eqs.  
\Eqref{Eq:SelfConsistency21} and \Eqref{Eq:SelfConsistency22} can be 
solved using an expansion by small parameters $\gamma/T_{c0} \ll 1$ and 
$\lambda_J \ll 1$. The first step is to find the gap amplitudes $\Delta_+$ 
and $\Delta_-$ in $s_{++}$ and $s_{\pm}$ states respectively with the 
accuracy up to linear order in $\gamma$ and $\lambda_J$. For the $s_{++}$ 
state we can use the exact expression 
$g=\omega_n/\sqrt{\omega_n^2+\Delta_{+}^2}$. For the $s_\pm$ 
one, we find linear corrections for the propagator using the equation 
$(\omega_n+2\gamma g)^2(1-g^2)=\Delta_-^2g^2$, which in general does not 
have an analytical solution. In this way we obtain that the $s_{++}$ and 
$s_\pm$ gap amplitudes satisfy 
\begin{align} 
 & 2\pi T \sum_{n=0}^{N_d}\frac{1}{(\omega_n^2+\Delta^2_+)^{1/2}} = 
 	\frac{1}{\lambda-\lambda_J}
 \,, \label{Eq:SelfConsistencySpp} \\ 
 & 2\pi T \sum_{n=0}^{N_d} \left[ \frac{1}{(\omega_n^2+\Delta^2_-)^{1/2}} 
 - \frac{2\gamma \omega^2_n}{(\omega_n^2+\Delta^2_-)^{2}} \right]
 = \frac{1}{\lambda+\lambda_J} 
 \,.  \label{Eq:SelfConsistencySpm}     
\end{align}
To find the boundaries of the $s+is$ domain we subtract 
Eqs.~\Eqref{Eq:SelfConsistency21} and \Eqref{Eq:SelfConsistency22} 
from each other to obtain 
\begin{align}
 & 2\pi T \sum_{n=0}^{N_d} \left[\frac{\gamma_+}{(\omega_n^2+\Delta^2_+)} 
 - \frac{2\gamma^2_+}{(\omega_n^2+\Delta^2_+)^{3/2}}\right]
  = \frac{\lambda_J}{\lambda^2}
  \,,  \label{Eq:TransSpp} \\
 & 2\pi T \sum_{n=0}^{N_d} \left[\frac{\gamma_-}{(\omega_n^2+\Delta^2_-)} 
 +\frac{2\gamma^2_- (\Delta^2_- - \omega_n^2) }{(\omega_n^2+\Delta^2_-)^{5/2}}\right] 
 = \frac{\lambda_J}{\lambda^2} 
 \,,  \label{Eq:TransSpm} 
 \end{align}    
where the Eqs.~\Eqref{Eq:TransSpp} and \Eqref{Eq:TransSpm} yield the 
implicit equations  $\gamma_+=\gamma_+(T)$ and $\gamma_- = \gamma_-(T)$ 
describing the $s_{++}/s+is$ and $s_{\pm}/s+is$ transition correspondingly.    
Here we need to take into account the terms up to the second order in 
$\gamma$ since in the linear order the transition lines coincide. 

The largest width for the $s+is$ domain occurs at low temperatures 
$T\ll T_{c}$ when we can substitute summation by Matsubara frequencies 
$2\pi T\sum_n = \int d\omega $. Hence, from Eqs.~\Eqref{Eq:SelfConsistencySpp}
and \Eqref{Eq:SelfConsistencySpm}, we obtain the relation between the gap 
functions in $s_{++}$ and $s_\pm$ phases: 
\begin{equation}\label{Eq:DpmRatio}
 \ln\left(\frac{\Delta_+}{\Delta_-}\right) = \frac{\pi}{2}\frac{\gamma_c}{\Delta_-} 
 - \frac{2\lambda_J}{\lambda^2}.
\end{equation}  
To the first order in small parameters, Eq.~\Eqref{Eq:DpmRatio} can be 
reduced to give the difference $\delta\Delta=\Delta_--\Delta_+=\gamma\pi/2$.
The relation \Eqref{Eq:DpmRatio} holds everywhere in the $s_\pm$ region, 
where it gives the dependence of gap function $\Delta_-=\Delta_-(\gamma)$, 
while the gap does not depend on $\gamma$ in the $s_{++}$ region. Within the 
same approximation, the Eqs.~\Eqref{Eq:TransSpp} and \Eqref{Eq:TransSpm} yield
 \begin{align} 
 & \frac{\pi}{2}\frac{\gamma_+}{\Delta_+} 
  - 2 \frac{\gamma^2_+}{\Delta_+^2} = \frac{\lambda_J}{\lambda^2} 
 \,, \label{Eq:TransSpp1} \\
 & \frac{\pi}{2}\frac{\gamma_-}{\Delta_-} 
  + \frac{2}{3} \frac{\gamma^2_-}{\Delta_-^2} 
  = \frac{\lambda_J}{\lambda^2}
 \,. \label{Eq:TransSpm1}
\end{align}
 
From the above argument, we conclude that the $s+is$ domain is located in 
the vicinity of $\gamma_c= 2\Delta_- \lambda_J /(\pi \lambda^2) $, and the 
width of the $s+is$ domain $\delta\gamma = \gamma_+ - \gamma_-$ is much 
smaller, $\delta\gamma \ll \gamma_c$. This can be found by combining 
Eqs.~\Eqref{Eq:TransSpp1} and \Eqref{Eq:TransSpm1}, which yield 
$\delta\gamma = (0.4/\pi) \gamma^2_c/\Delta_+$, so that 
\begin{equation}\label{Eq:WidthSiS}
\delta\gamma =  \frac{\Delta_+}{20}\frac{\lambda_J^2}{\lambda^4}  .
\end{equation}    
For the parameters used in \Figref{Fig:Transition-zoom} the estimate 
\Eqref{Eq:WidthSiS} yields $\delta\Gamma = 2\delta\gamma \propto 10^{-3}$, 
which coincides by the order of magnitude with numerically found values.    
In general Eq.~\Eqref{Eq:WidthSiS} implies that the $s+is$ domain is 
generically narrow  for this model of a two-band superconductor with 
impurities since its width is determined by the parameter 
$\lambda_J^2/\lambda^4$  which is small according to the restriction 
\Eqref{Eq:TcRestriction}. 
   
The characteristic shape of the $s+is$ domain, in the plots of 
\Figref{Fig:Diagram} can be understood considering one of the transition 
lines, namely between $s_{++}$ and $s+is$ domains. With good accuracy we 
can use the first-order-in-$\gamma$ approximation in 
Eq.~\Eqref{Eq:TransSpp}, which yields
 \begin{equation} \label{Eq:TransSpp1order}
  \frac{\gamma_+}{2\pi T} \psi^{\prime} 
  \left(\frac{\Delta_+}{2\pi T}+ \frac{1}{2}\right)
  =  - \frac{\lambda_J}{\lambda^2} \,,
\end{equation}
where $\psi(s)$ is a digamma function satisfying the relation
$\psi(s) - \psi(1/2) =\sum_{n=0}^{\infty} [ 1/(n+1/2) - 1/(n+s)]$. 
Given that the temperature dependence of the gap $\Delta_+=\Delta_+(T)$ 
is determined by the usual gap equation \Eqref{Eq:SelfConsistencySpp}, 
one can find that the Eq.~\Eqref{Eq:TransSpp1order} yields only a weak 
temperature variation of $\gamma_+(T)$ which is consistent with the 
$s+is$ domain being elongated, almost parallel to the $T$ axis as can 
be seen in the first column of \Figref{Fig:Diagram}.

\subsection{Effect of the band disparity}

We have previously shown that, for exactly degenerate bands, the $s+is$ 
state extends all the way up to the critical temperature near the $s_\pm$ 
to $s_{++}$ transition. As can be seen from the numerical plots in 
\Figref{Fig:Diagram}, for the case of finite band disparity, the $s+is$ 
dome is disconnected from the $T_{c}(\Gamma)$ curve. It is instead replaced 
by a crossover line near the critical temperature. To describe this effect 
analytically, let us derive the equation describing $s+is$ transition lines, 
assuming again that the condition \Eqref{Eq:TcRestriction} is satisfied, 
so that the interband scattering amplitudes are small. Here, we are not 
interested in the width of the $s+is$ region, so that we implement 
first-order expansion in $\gamma_{12}$ to obtain from 
Eqs.~\Eqref{Eq:Eilenberger:1} and \Eqref{Eq:Eilenberger:2}
\begin{equation} \label{Eq:f1LinearGamma}
  f_1= \Delta_1\frac{g_1}{\omega_n} 
  + (\Delta_2-\Delta_1)\frac{\gamma_{12}g_1g_2}{\omega_n^2} \,. 
\end{equation}
A similar expression for $f_2$ is given by the interchange 
$1 \leftrightarrow 2$ in Eq.~\Eqref{Eq:f1LinearGamma}. Substituting 
\Eqref{Eq:f1LinearGamma} into the self-consistency equation 
Eq.~\Eqref{Eq:SelfConsistency1}, we find that on both $s_{\pm}/s+is$ 
and $s_{++}/s+is$ transition lines, the following condition is 
satisfied 
\begin{equation}\label{Eq:TransitionLines}
  2\pi T\sum_{n=0}^{\infty} \frac{g_1-g_2}{\omega_n} 
  = \frac{1}{\lambda_{11}} - \frac{1}{\lambda_{22}} \,,
\end{equation}
where we took into account that 
$\lambda_{12}\lambda_{21} \ll \lambda_{11}\lambda_{22}$.
 
The relation \Eqref{Eq:TransitionLines} implies several properties of 
the $s+is$ transition lines. First it is clear that the $s+is$ domain does 
not reach $T_c$, since the condition \Eqref{Eq:TransitionLines} is not 
satisfied near the critical temperature where $g_1=g_2=1$. Besides that, 
near $T_c$ we can put $g_2=1$ and $g_1=1-\Delta_1^2/2\omega_n^2$, to 
rewrite the condition \Eqref{Eq:TransitionLines} in the simpler form 
\begin{equation}\label{Eq:TransitionLines1}
 \Delta_1^2 = \frac{8\pi^2 T^{2}_c}{7\zeta(3)}  
 \frac{ \lambda_{11}-\lambda_{22} }{ \lambda_{11}\lambda_{22}},
\end{equation} 
where $\zeta(3)=1.2$ is the Riemann zeta-function. Provided that 
$\Delta_1^2 \propto T_c(T_c-T)$, it is clear that the $s+is$ states can 
extend only up to the threshold temperature 
$T^* =  T_c [1+  \alpha( \lambda_{11}^{-1} - \lambda_{22}^{-1} )] $, 
where $\alpha\sim 1$. When increasing band disparity, the temperature 
$T^*$ goes down, which agrees with the numerical results displayed in 
the first column of the \Figref{Fig:Diagram}.
  
The second important consequence of Eq.~\Eqref{Eq:TransitionLines1} 
follows from the fact that the gap amplitudes $|\Delta_1|$ are different 
on the $s_{\pm}$ and $s_{++}$ sides. Therefore the threshold temperature $T^*$ 
is different for the $s_{\pm}/s+is$ and $s_{++}/s+is$ transition lines. 
Consequently the phase diagram featuring the $s+is$ state is in general 
not only quantitatively but also qualitatively different from the plot 
given in \cite{Stanev.Koshelev:14}. Namely the $s_\pm$ to $s_{++}$ crossover 
line is not, in general, attached to the summit of the $s+is$ dome. Rather, 
it can attach to an arbitrary point of the line of second order phase transition 
that separates the $s+is$ state. This can for example be seen for a zoomed 
in diagram corresponding to panel (A3) of \Figref{Fig:Diagram}.

\subsection{ Domain of \texorpdfstring{$s+is$}{sis} state inside the 
			\texorpdfstring{$s_\pm$}{s+-} phase} 
\label{SubSec:sisInside}

\begin{figure}[!b]
\hbox to \linewidth{ \hss
\includegraphics[width=0.9\linewidth]{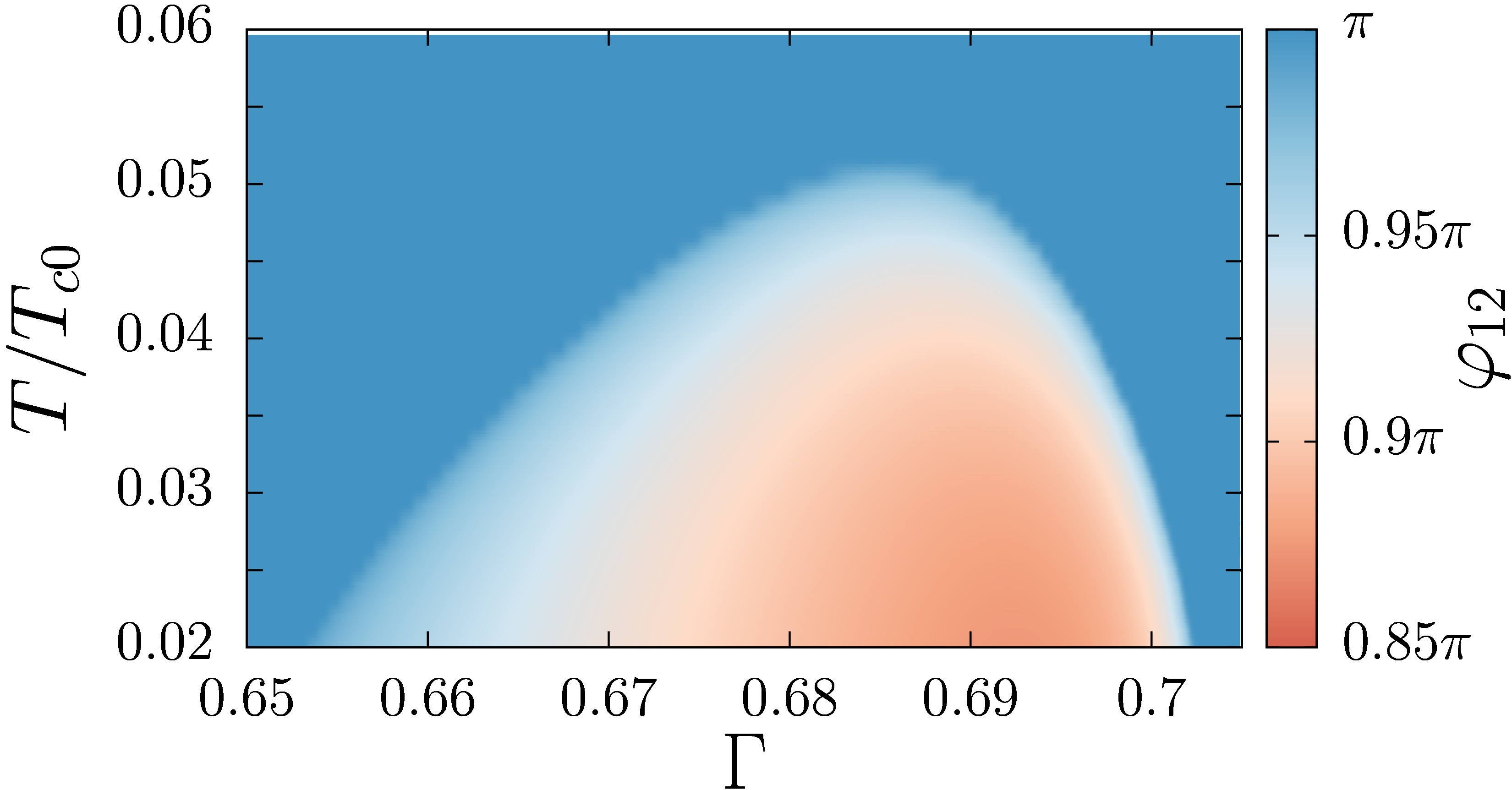}
\hss}
\caption{
(Color online) -- 
This displays a close-up view of the ``isolated dome'' in panel (C2) 
of \Figref{Fig:Diagram}. This demonstrates that the $s+is$ state 
can actually be disconnected from transition between $s_\pm$ and 
$s_{++}$ states.
}
\label{Fig:Dome}
\end{figure}

As previously emphasized, in general, the crossover line does not attach 
to the top of the $s+is$ dome. We find that it rather attaches to a 
different point belonging to the $s+is/s_{++}$ transition line.
This implies that, the $s+is$ state also can occur away from the 
$s_\pm$ to $s_{++}$ crossover. More precisely, as can be seen in 
\Figref{Fig:Dome} (corresponding to a zoom on the panel (C2) of 
\Figref{Fig:Diagram}), the $s+is$ state can show up as a small 
``isolated dome'' inside the $s_\pm$ region. 
Such an isolated dome can occur for rather important band disparity 
and intermediate values of the interband impurity scattering. This 
effect can be understood as follows. When there is band disparity and 
impurities, the crossover line attaches below the summit of the $s+is$ 
dome. Moreover, the $s+is$ region is generically pushed to lower 
temperatures when increasing band disparity. It can thus occur that 
at a certain level of disparity the temperature where the crossover 
line attaches to the dome goes to zero and then the crossover line 
detaches from the dome, which means that the $s+is$ dome becomes isolated.

\section{Order of the phase transitions in the Ginzburg-Landau model} 
\label{Sec:GL}
 
\begin{figure*}[!htb]
\hbox to \linewidth{ \hss
\includegraphics[width=.8\linewidth]{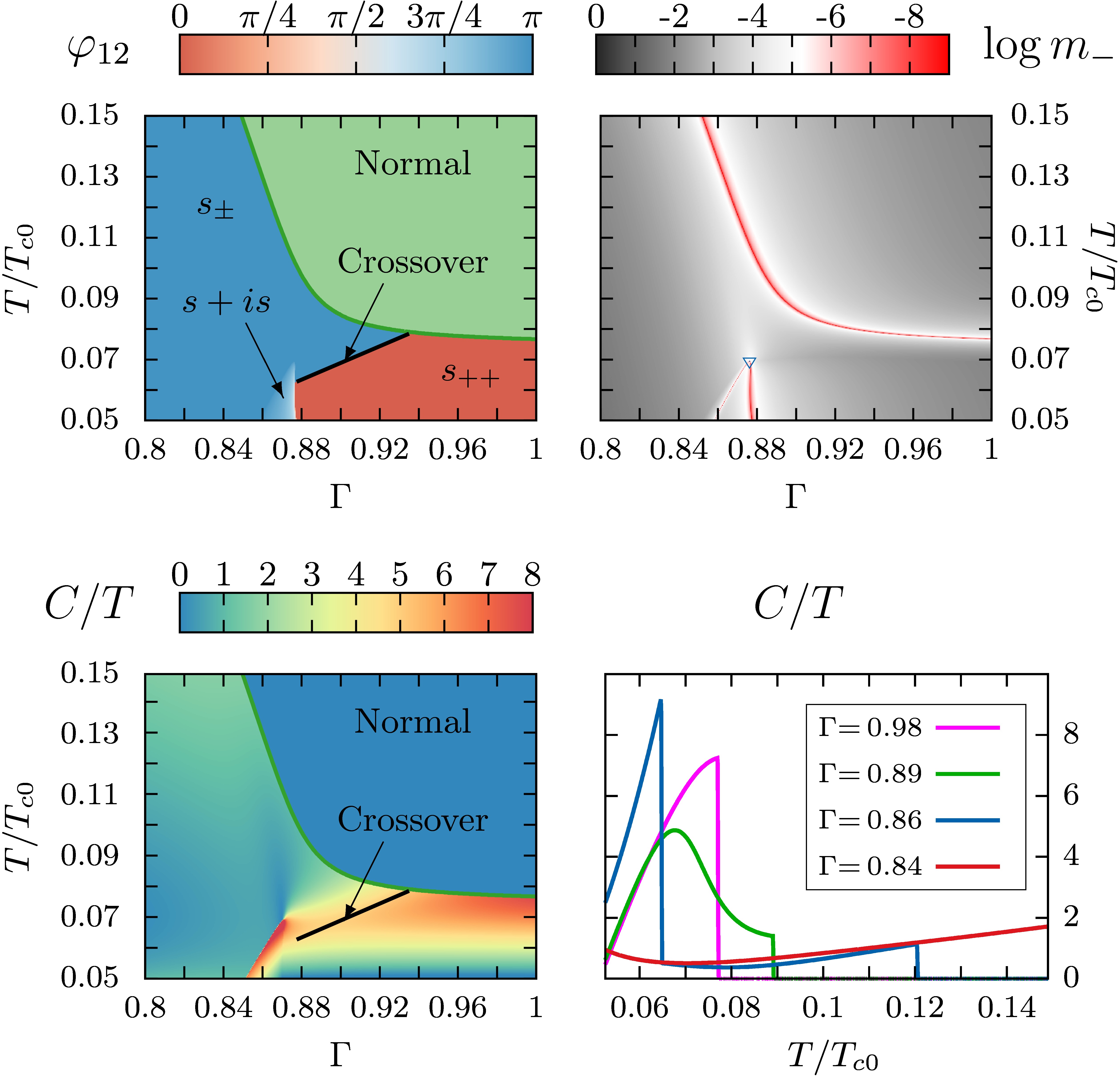}
\hss}
\caption{
(Color online) -- 
Top-left panel shows the phase diagram of two-band superconductors 
with interband impurity scattering for parameters corresponding to 
the panel (A3) in \Figref{Fig:Diagram}. That is, for the values of 
intraband coupling parameters are $\lambda_{11}=0.3$, 
$\lambda_{22}=0.29$ and $\lambda_{12}=\lambda_{21}=-0.1$. 
The top-right panel displays the smallest eigenvalue of the Hessian 
matrix of the Ginzburg-Landau free energy. Clearly, it vanishes 
at $T_c$ signaling that the system has a divergent coherence length
at these temperatures and thus the second order phase transition 
between superconducting and normal state. Moreover, the $s+is$ phase 
is surrounded by another line with vanishing smallest eigenvalue 
signalling the flatness of the potential and thus an additional second 
order transition from $s_\pm/s_{++}$ to $s+is$ states. 
Note that for the $s_\pm/s_{++}$ crossover, all eigenvalues remain 
finite and the second gap vanishes.
The triangle shows the summit of the $s+is$ dome, where a second 
eigenvalue vanishes as well.
The bottom-left panel shows the corresponding specific heat, while 
the bottom-right panel displays few one-dimensional cross sections 
of that plot, corresponding to vertical scans (at a given impurity).
There are clearly two jumps when the vertical lines intersects both 
the superconducting  phase transition as well as the phase transition 
from $s_\pm$ to $s+is$ state.
}
\label{Fig:DiagramGL}
\end{figure*}

As discussed in the first part of the paper, the $s+is$, state although 
realized, can be extremely challenging to observe due to the narrowness 
of the interval in parameter space where this state can exist. Nonetheless, 
it is a relevant question to study the order of the phase transition 
to the $s+is$ state, that is, if the phase transition is of the second
order, which guarantees that there appears a divergent coherence length 
that, in some range of parameters, should exceed the magnetic field 
penetration length as well as other coherence lengths 
\cite{Carlstrom.Garaud.ea:11a,Garaud.Silaev.ea:16a}. Regimes with some 
coherence lengths larger, and some smaller than the magnetic field 
penetration length feature attractive inter-vortex forces that, under 
some circumstances, may be responsible for formation of vortex 
clusters (that kind of regimes was termed type-1.5 superconductivity in 
Ref.~\cite{Moshchalkov.Menghini.ea:09}). Therefore due to the divergent 
behaviour of one of the coherence lengths at the phase transition there 
may be a range of this state with anomalous magnetic and transport 
properties.

To determine the order of the phase transition at the mean-field level
we derive a Ginzburg-Landau free energy functional from the microscopic 
equations. Here we implement the standard multiband expansion in two 
small gaps $\Delta_k=|\Delta_k|\Exp{i\varphi_k}$ in the dirty case 
(note that the multiband expansions in general are based on assumptions 
of several small parameters \cite{Tilley:64} that are not related to 
broken symmetries, not to be confused with the simplest expansion in 
single small parameter $\tau=1-T/T_c$). Justification and validity 
conditions of multiband expansions of this kind in several small 
parameters were discussed in detail in the clean $s$-wave case 
\cite{Silaev.Babaev:12}. The potential terms in such an expansion 
read as:
\begin{align}
 \F&=\sum_{k=1}^2\Big\{a_{kk}|\Delta_k|^2+\frac{b_{kk}}{2} |\Delta_k|^4 \Big\} 
 \nonumber \\
 &+2\Big(a_{12}+c_{11}|\Delta_1|^2+c_{22}|\Delta_2|^2\Big)
 |\Delta_1||\Delta_2|\cos\varphi_{12}\nonumber \\
 &+\Big(b_{12}+c_{12}\cos2\varphi_{12}\Big)|\Delta_1|^2|\Delta_2|^2
 \,. \label{Eq:FreeEnergy}
\end{align}
There, the coefficients $a_{kk^\prime}$, $b_{kk^\prime}$, and $c_{kk^\prime}$ 
can be calculated from the inputs $\lambda_{kk^\prime}$, $T$, and 
$\Gamma$ of the microscopic self-consistent equations 
\cite{Stanev.Koshelev:14}. 

We investigate the state properties of the Ginzburg-Landau theory 
by minimizing the free energy \Eqref{Eq:FreeEnergy} with respect 
to the densities $|\Delta_k|$ and the relative phase 
$\varphi_{12}=\varphi_2-\varphi_1$. The relative phase is 
given by the equation $\delta\F/\delta\varphi_{12}=0$:
\begin{align}
 \Big(a_{12}&+c_{11}|\Delta_1|^2+c_{22}|\Delta_2|^2\Big)
 |\Delta_1||\Delta_2|\sin\varphi_{12} \nonumber \\
 &+c_{12}|\Delta_1|^2|\Delta_2|^2\sin2\varphi_{12}=0 
 \,. \label{Eq:PhaseDifference}
\end{align}
This has different solutions in the different  states:
\begin{align}
 s_{\pm}&:~\varphi_{12}=\pi\,,~~~~ s_{++}:~\varphi_{12}=0
 \,, \label{Eq:GSPhaseDifference} \\
 s+is&:~\varphi_{12}=
 \pm\arccos\left(-\frac{a_{12}+c_{11}|\Delta_1|^2+c_{22}|\Delta_2|^2}
 {2c_{12} |\Delta_1||\Delta_2|}\right) 	\,.\nonumber
\end{align}
Note here that the values of the gaps are determined by the other 
equations $\frac{\delta\F}{\delta|\Delta_k|}=0$. \Figref{Fig:DiagramGL} 
shows an example of such a calculation applied to the regime displayed 
in panel (A3) of \Figref{Fig:Diagram}. In the area where the two-band 
Ginzburg-Landau expansion in small gaps is formally justified, the 
phase diagram matches with that obtained within the microscopic theory 
of Eqs.~\Eqref{Eq:Eilenberger:1}--\Eqref{Eq:SelfConsistency1}). 

Now we focus on the properties of the transition lines between the 
various phases. For a given state to be stable, all the eigenvalues 
of the stability (Hessian) matrix $\hat{\cal H}$ must be positive. 
The Hessian matrix reads as:
\begin{equation}
 \hat {\cal H} = \left(
  \begin{array}{ccc}
    \frac{\partial^2 \F}{\partial^2 \Delta_1} 
    & \frac{\partial^2 \F}{\partial \Delta_1\partial\Delta_2} 
    & \frac{\partial^2 \F}{\partial \Delta_1\partial\varphi_{12}} \\
    \frac{\partial^2 \F}{\partial\Delta_1\partial\Delta_2} 
    & \frac{\partial^2 \F}{\partial^2\Delta_2} 
    & \frac{\partial^2 \F}{\partial \Delta_2\partial\varphi_{12}} \\
    \frac{\partial^2 \F}{\partial\Delta_1\partial\varphi_{12}} 
    & \frac{\partial^2 \F}{\partial \Delta_2\partial\varphi_{12}} 
    & \frac{\partial^2 \F}{\partial^2\varphi_{12}} \\
  \end{array}
\right)
 \end{equation}
and ${\cal H}_{ab}$, where $a,b=1,2,\varphi$ denotes the entry 
of the Hessian that relates to the variations with respect to 
$|\Delta_1|$, $|\Delta_2|$ and $\varphi_{12}$ respectively. 
Second order phase transitions are associated with a divergent coherence 
length, and, correspondingly, there should be flatness of the potential 
in some direction of parameter space. Mathematically, this can be 
characterized by the vanishing of the smallest eigenvalue of the 
corresponding Hessian matrix. 

By definition, the $s+is$ state is that where both $|\Delta_1|$ 
and $|\Delta_2|$ are nonzero and for which $\varphi_{12}\neq0,\pi$  
is given by the equation $\delta\F/\delta\varphi_{12}=0$. 
Both in the $s_{\pm}$ and $s_{++}$ state ${\cal H}_{i\varphi}=0$ 
($i=1,2$). And thus the contribution of the relative phases to 
the stability (Hessian) matrix originates only in 
${\cal H}_{\varphi\varphi}$. The stability of the $s_{\pm}$ and 
$s_{++}$ states thus requires that 
${\cal H}_{\varphi\varphi}\big|_{s_\pm,s_{++}}>0$. The transition 
line to the $s+is$ is thus given by the condition that 
${\cal H}_{\varphi\varphi}\big|_{s_\pm,s_{++}}=0$. Indeed, this 
is the point where $s_{\pm}$ and $s_{++}$ states become unstable 
and $s+is$ becomes stable. This gives an additional condition, for the 
transition line:
\begin{align}
\Big(a_{12}&+c_{11}|\Delta_1|^2+c_{22}|\Delta_2|^2\Big)
|\Delta_1||\Delta_2|\cos\varphi_{12} \nonumber \\
&+2c_{12}|\Delta_1|^2|\Delta_2|^2\cos2\varphi_{12}=0 
\,.	\label{Eq:Transition}
\end{align}

The right panel in \Figref{Fig:DiagramGL} displays the smallest 
eigenvalue of the Hessian matrix of the Ginzburg-Landau free energy. 
Clearly, it vanishes at $T_c$ which means that there is a divergent 
coherence length and thus the standard, at mean-field level,  second 
order phase transition between  superconducting and normal state. 
Moreover, the $s+is$ phase is surrounded by another line with vanishing 
smallest eigenvalue signaling the flatness of potential, and thus a 
second-order transition and a divergent coherence length at the transition 
to the $s+is$ state. The corresponding behavior of the specific heat 
$C=-TdS/dT$ following from Eq.~\Eqref{Eq:FreeEnergy} is displayed in the 
bottom line of \Figref{Fig:DiagramGL}. By contrast the total density does 
not have a strong feature at the phase transition (see \Figref{Fig:Density}). 
Therefore no dramatic variation of magnetic field penetration 
length is expected.

\section{Conclusion} \label{Sec:Conclusion}
 
We discussed the phase diagram of dirty two-band superconductors. 
Our main interest was to examine the possible occurrence of the $s+is$ 
superconducting state in the  mean-field  model. We find that the 
state exists and appears under more general circumstances than previously 
discussed: namely it is not necessarily connected to the crossover 
from the $s_\pm$ to $s_{++}$ state but can arise inside the $s_\pm$ phase. 
However, we demonstrate that the domain of the $s+is$ state is extremely 
small in this model and on a large-scale plots  almost shrinks to a 
line. For all practical purposes that makes it unobservable in 
experiments in materials with this microscopic physics.  We also 
establish that the phase transitions to the $s+is$ state are second 
order on the mean field level. This implies  that near the phase transition
there is a divergent length scale associated with the order-parameters 
variations, as well as a related softening of dynamical modes:
such as Leggett's \cite{Lin.Hu:12} and ``phase-density mixed" collective 
modes \cite{Carlstrom.Garaud.ea:11a,Stanev:12,Maiti.Chubukov:13,
Marciani.Fanfarillo.ea:13}. We emphasise that our ``non-observability" 
results here apply only for the weak-coupling models of two-band 
superconductivity but do not preclude formation of larger areas of 
$s+is$ states by other mechanisms such as three-band systems with 
frustrated intercomponent interaction.

\begin{acknowledgments}
This work was supported by the Swedish Research Council Grant  
No.~642-2013-7837 and by the Goran Gustafsson Foundation.   
M. S. was supported in part by the Academy of Finland.
The computations were performed on resources provided by the 
Swedish National Infrastructure for Computing at National 
Supercomputer Center at Link\"oping, Sweden.
\end{acknowledgments} 


\appendix
\section{Numerical procedure}

Given a set of microscopic parameters (temperature $T$, interband 
scattering amplitude $\Gamma$, and couplings $\lambda_{kk^\prime}$), 
the state is found by solving numerically the Eilenberger equations 
\Eqref{Eq:Eilenberger:1} and \Eqref{Eq:Eilenberger:2}, under the 
condition that $|f_k|^2 + g_k^2 =1$, together with the self-consistency 
equation \Eqref{Eq:SelfConsistency1}, following the procedure described 
below.

First, note that using Eqs.~\Eqref{Eq:Eilenberger:1} and
\Eqref{Eq:Eilenberger:2}, the anomalous Green's functions can be 
expressed in terms of the normal Green's functions, gaps, etc, as:
\begin{align}
\left(\begin{array}{c} f_1 \\ f_2\end{array}\right)&=\frac{1}{w}
\left(\begin{array}{cc}
  g_1(\omega_n+\gamma_{21}g_1) 		 	& \gamma_{12}g_1g_2 \\
  \gamma_{21}g_1g_2 	& g_2(\omega_n+\gamma_{12}g_2) \\
\end{array}\right)
\left(\begin{array}{c} \Delta_1 \\ \Delta_2\end{array}\right) 
\nonumber \\
~~\text{with}&~~w=\omega_n(\omega_n+\gamma_{12}g_2+\gamma_{21}g_1)
\,.\label{Eq:Numerics:AnomalousGF}
\end{align}
This allows one to find expressions for $|f_k|^2$, which once substituted 
into the normalization condition yields the system of nonlinear 
equations for the normal Green's functions $g_k$, as: 
\begin{align}
w^2&(g_1^2-1)+g_1^2\big[ (\omega_n+\gamma_{21}g_1)\Re(\Delta_1)
	+\gamma_{12}g_2 \Re(\Delta_2) \big]^2 
    \nonumber \\
+&g_1^2\big[ (\omega_n+\gamma_{21}g_1)\Im(\Delta_1)
	+\gamma_{12}g_2 \Im(\Delta_2) \big]^2 = 0 \,,
	\label{Eq:Numerics:NormalGF:1} \\
w^2&(g_2^2-1)+g_2^2\big[ (\omega_n+\gamma_{12}g_2)\Re(\Delta_2)
	+\gamma_{21}g_1 \Re(\Delta_1) \big]^2 
	\nonumber \\
+&g_2^2\big[ (\omega_n+\gamma_{12}g_2)\Im(\Delta_2)
	+\gamma_{21}g_1 \Im(\Delta_1) \big]^2 = 0 \,.
\label{Eq:Numerics:NormalGF:2}
\end{align}
We choose an optimization method to find $g_{1,2}$ from Eqs.~
\Eqref{Eq:Numerics:NormalGF:1} and \Eqref{Eq:Numerics:NormalGF:2} 
based on an objective function $F$ given by 
\begin{equation}\label{Eq:Numerics:NormalGF:Obj}
F(g_1,g_2)=   \big[\mathrm{Eq}.\,(\ref{Eq:Numerics:NormalGF:1}) \big]^2
	+\big[\mathrm{Eq}.\,(\ref{Eq:Numerics:NormalGF:2}) \big]^2				\,.
\end{equation}
We use a Covariance Matrix Adaptation Evolution Strategy (CMA-ES) 
\cite{Hansen:06}, which is a stochastic numerical optimization method 
for non-linear or non-convex problems. As compared to other algorithms, 
this can be rather suboptimal, but since it is a stochastic, gradient-free 
method, the solution is guaranteed to be independent of any initial guess. 

The anomalous Green's functions can be reconstructed, given the 
solutions $g_k(\omega_n)$ of Eqs.~\Eqref{Eq:Numerics:NormalGF:1} and 
\Eqref{Eq:Numerics:NormalGF:2}, using Eq.~\Eqref{Eq:Numerics:AnomalousGF}. 
Finally, the gaps are constructed using the self-consistency equation 
\Eqref{Eq:SelfConsistency1}. This procedure is iterated via a fixed-point 
method until the gaps converge according to the criterion that 
$\sqrt{\sum_k\left|\Delta_k^{\mbox{\tiny new}}
-\Delta_k^{\mbox{\tiny old}}\right|^2}<10^{-7}$.

%

\end{document}